\documentclass[12pt,preprint]{aastex}

\usepackage{epsfig}
\usepackage{amsmath}
\usepackage{subfigure}
\usepackage{natbib}
\usepackage{multirow}
\usepackage{txfonts}
\begin{document}

\title{Discovery of SiCSi in IRC\,+10216: A missing link between gas and dust carriers of Si$-$C bonds}
\shorttitle{Discovery of SiCSi}
\shortauthors{Cernicharo et al.}

\author
{
J. Cernicharo\altaffilmark{1}, 
M. C. McCarthy\altaffilmark{2},
C. A. Gottlieb\altaffilmark{2},
M. Ag\'undez\altaffilmark{1},
L. Velilla Prieto\altaffilmark{1},
J. H. Baraban\altaffilmark{3},
P. B. Changala\altaffilmark{4},
M. Gu\'elin\altaffilmark{5},
C. Kahane\altaffilmark{6},
M. A. Martin-Drumel\altaffilmark{2},
N. A. Patel\altaffilmark{2},
N. J. Reilly\altaffilmark{2,7},
J. F. Stanton\altaffilmark{8},
G. Quintana-Lacaci\altaffilmark{1},
S. Thorwirth\altaffilmark{9},
K. H. Young\altaffilmark{2}
}

\altaffiltext{1}{Group of Molecular Astrophysics. ICMM. CSIC. C/Sor Juana In\'es de La Cruz N3. E-28049, Madrid. Spain}
\altaffiltext{2}{Harvard-Smithsonian Center for Astrophysics, Cambridge, MA 02138, and School of Engineering \& Applied Sciences, Harvard University, Cambridge, MA 02138}
\altaffiltext{3}{Department of Chemistry and Biochemistry, University of Colorado, Boulder, CO 80309}
\altaffiltext{4}{JILA, National Institute of Standards and Technology and University of Colorado, and Department of Physics, University of Colorado, Boulder, CO 80309}
\altaffiltext{5}{Institut de Radioastronomie Millim\'etrique, 300 rue de la Piscine, F-38406, St-Martin d'H\`eres, France}
\altaffiltext{6}{Université Grenoble Alpes, IPAG, F-38000 Grenoble, France; CNRS, IPAG, F-38000 Grenoble, France}
\altaffiltext{7}{Present address: Department of Chemistry, Marquette University, Milwaukee, WI 53233}
\altaffiltext{8}{Institute for Theoretical Chemistry, Department of Chemistry, The University of Texas at Austin, Austin, TX 78712}
\altaffiltext{9}{I. Physikalisches Institut, Universit\"at zu K\"oln, Z\"ulpicher Str. 77, 50937 K\"oln, Germany}

\date{To be published in the ApJ Letters; received April 21st, 2015; accepted May 06, 2015}

\begin{abstract}
We report the discovery in space of a disilicon species, SiCSi,
from observations between 80 and 350 GHz with the IRAM\footnote[10]{This work was based on observations 
carried out with the IRAM 30-meter telescope. IRAM is 
supported by INSU/CNRS (France), MPG (Germany) and IGN (Spain)}
 30m radio telescope.
Owing to the close coordination between 
laboratory experiments and astrophysics, 112 lines have now been detected in the
carbon-rich star CW\,Leo. The derived frequencies yield improved
rotational and centrifugal distortion constants up to sixth order. From the line profiles and 
interferometric maps with the Submillimeter Array\footnote[11]{The Submillimeter Array is a joint project between the Smithsonian
Astrophysical Observatory and the Academia Sinica Institute of Astronomy and Astrophysics, and is funded by the Smithsonian Institution and the
Academia Sinica.}, the bulk of the SiCSi emission arises from a 
region of 6$''$ in radius.
The derived abundance is comparable to that of SiC$_2$. As expected from 
chemical equilibrium calculations, SiCSi and SiC$_2$ are the most abundant species harboring a
Si$-$C bond in the dust formation zone and certainly both play a key role in the formation of SiC dust grains.
\end{abstract}

\section{Introduction}

Interstellar dust grains are synthesized in two main types of sources: the inner winds of asymptotic giant
 branch (AGB) stars and the ejecta of massive stars, mainly supernovae (SNe).
Dust grains are formed from molecular seeds. In AGB stars, molecules such as TiO, VO, ZrO, C$_2$, CN, and 
C$_3$, among others, are known to be present in their photospheres since the beginning of the 20th century. 
Because of the high stability of the CO molecule, depending on whether the C/O ratio is $>$1 (C stars) 
or $<$1 (M stars), the gas becomes either carbon- or oxygen-rich after CO reaches its equilibrium 
abundance, leading to a different chemistry and therefore to different dust formation schemes. 
In SNe, different types of dust can be formed depending on the degree of enrichment in heavy elements. 
The dust formation efficiency in SNe is still a matter of debate. 
Recent studies with the \emph{Herschel} Space Telescope point to much larger dust masses than previously 
thought \citep{Matsuura2011,Matsuura2015}.

Dust formation can be simplified as a two-step process: formation of nucleation seeds followed by grain 
growth 
through condensation of refractory molecules at high temperature and other less refractory species at 
larger distances from the star. There are, however, many mysteries in this picture of events, starting 
from the fundamental step of the formation of the nucleation seeds, which essentially are
refractory species \citep{Gail2010}. 
The presence of SiC grains in C-rich AGBs 
was confirmed by the detection of an emission band at $\sim$11.3 $\mu$m \citep{Treffers1974}. This band 
has been found towards a large number of C-rich stars with the IRAS and ISO satellites. However, the 
molecular precursors of SiC dust grains are still unknown. SiC molecules have been detected in the 
external shells ($\ge$300 R$_*$) of IRC\,+10216 \citep{Cernicharo1989,Patel2013}. In the inner layers, 
the most abundant molecule harbouring a Si$-$C bond is SiC$_2$  \citep{Thaddeus1984, Cernicharo2010}, 
whereas the most abundant Si-bearing species is SiS, which, together with SiO and SiC$_2$, 
account for a significant fraction of the available silicon \citep{Agundez2012}.

CW Leo, located $\simeq$130\,pc from us, is a Mira variable star with a period of 630-670 days and an 
amplitude of $\simeq$1 mag in the $K$ band \citep
{Menten2012}. It is one of the 
brightest infrared sources in the sky. Due to its close proximity, IRC\,+10216, the circumstellar envelope 
of CW Leo, has attracted many studies because it is exceptionally rich in molecular species. 
Half of the known interstellar species are observed in this C-rich envelope. The observed molecules range 
from CO, the main tracer of the cool molecular gas, and other diatomic and triatomic species 
\citep[see e.g.][]{Cernicharo2000,Cernicharo2010}, to molecules containing refractory elements 
\citep{Cernicharo1987}, and long carbon chain species C$_n$H and their anions \citep[][and references 
therein]{Cernicharo1996,Cernicharo2008,Guelin1997,McCarthy2006,Thaddeus2008}. Among the species detected 
in this source, the silicon-carbon species Si$_n$C$_m$ could play an important role as seeds of SiC dust 
grains. The simplest members of such a family, SiC \citep{Cernicharo1989}, 
and SiC$_2$ \citep{Thaddeus1984}, are known to be present in IRC\,+10216. However, 
Si$_2$C (hereafter SiCSi), which is predicted 
to be very abundant from chemical equilibrium calculations \citep{Tejero1991,Yasuda2012} and could play a key 
role in the formation of SiC dust grains, has not been found so far.

In this Letter we present the detection of SiCSi in IRC\,+10216 based on a close coordination 
between laboratory, astrophysics and radio astronomy. This synergy has resulted in
the detection of 112 lines of this new species using data from the IRAM 30-m radio telescope. Nine of 
these lines are also detected with the Submillimeter Array (SMA).

\section{Observations}

In the course of searches for new molecules we have covered a large fraction of the 3, 2, 1 and 
0.8 mm spectrum of IRC+10216 with a high sensitivity
using the 30-m IRAM radio telescope. In the 3 mm window, the data 
acquired during the last 30 years cover the 80-116 GHz domain with very high 
sensitivity (1-3 mK). Examples of these data can be found in 
\citet{Agundez2008,Agundez2014}, \citet{Cernicharo1996}, \citet{Cernicharo2007,Cernicharo2008}, 
and references therein. Most of the 2 mm data come from the line survey of IRC\,+10216 
carried out by \citet{Cernicharo2000}, complemented with additional data obtained during the 
search for specific molecular species \citep[see, e.g.][]{Guelin2004,Agundez2008,Agundez2012,Fonfria2006}. 
The sensitivity of the 2 mm observations varies between 1.5 and 10 mK. The 1 mm and 0.8 mm data come 
from observations carried out during the searches quoted above and a line survey between 290 and 355 
GHz carried out using the new EMIR receivers in 2010 and 2011
with a sensitivity of 2-5 mK, depending on the atmospheric transmission. 

For frequencies above 250 GHz the spectrometers were two autocorrelators with 2 MHz of spectral 
resolution and 4 GHz of bandwidth. For all other observations the spectral resolution was 1 MHz 
provided by filter banks or autocorrelators. 
The observing mode, in which we wobbled the secondary mirror by $\pm$90'' at a 
rate of 0.5 Hz, and the dry weather conditions (sky opacity at 225 GHz was below 0.1 in most 
observations) ensured flat baselines
and low system noise temperatures ($T_{\rm sys}$ $\simeq$ 100-400 K depending on the frequency).

\begin{figure}
\begin{center}
\includegraphics[angle=0,scale=0.65]{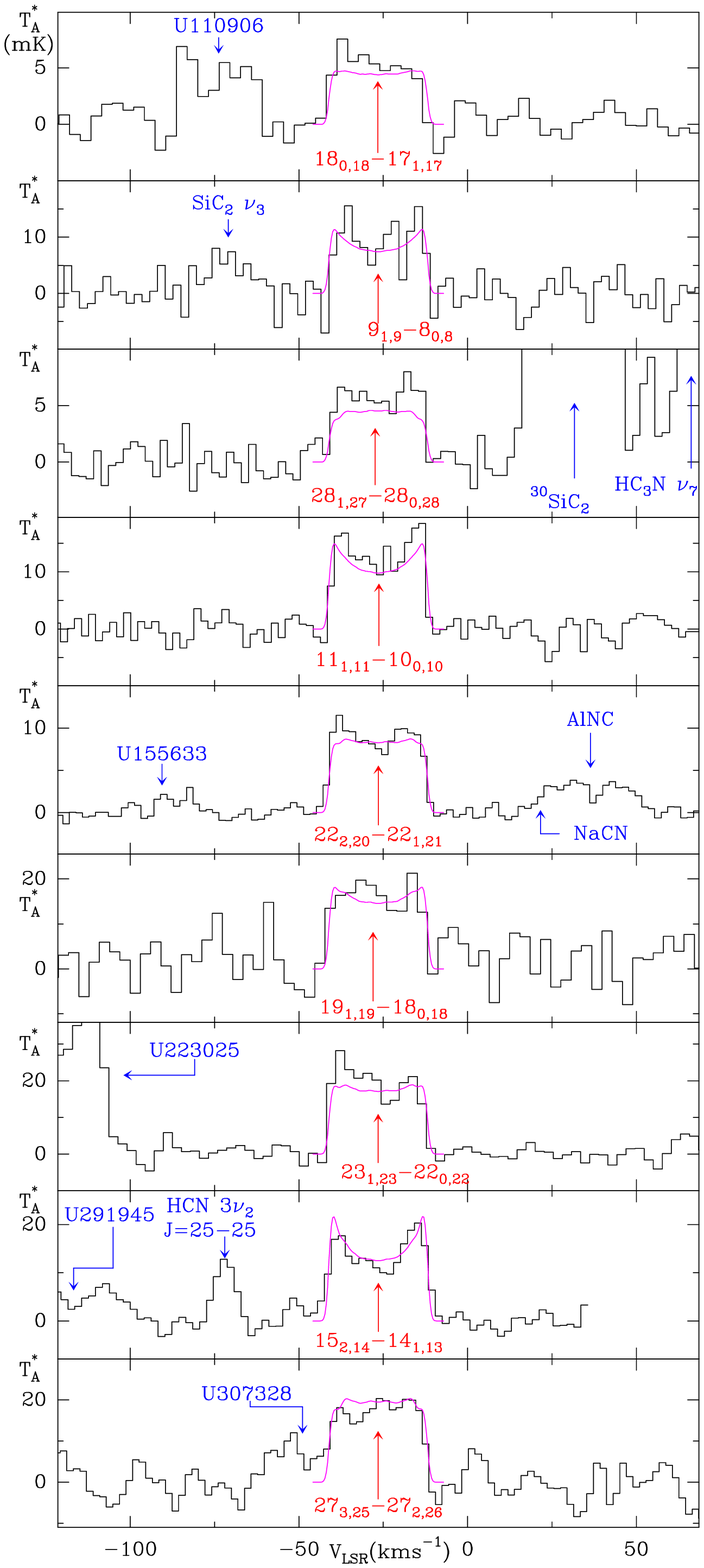}
\caption{Selected transitions of SiCSi among the 112 observed lines. Unidentified (U)
lines are indicated by their frequency in MHz.
The intensity scale is antenna temperature in mK.
The data are shown as histograms. 
Violet continuum lines show the predicted line profile from the best model
for the abundance of SiCSi (see text).}
\label{Fig1}
\end{center}
\end{figure}

The selected observing method, with the off position located at 180$''$ from the star, 
provides reference data free from emission for all molecular species but CO
\citep[see][]{Cernicharo2015}. The emission of all other 
molecular species is restricted to a region $\le$ 20-30$''$ \citep[see, e.g.,][]{Guelin1993}. 
The intensity scale, antenna temperature ($T_A^*$), was corrected for atmospheric absorption using the ATM 
package \citep{Cernicharo1985,Pardo2001}. The main beam antenna temperature can be obtained by 
dividing $T_A^*$ by the main beam efficiency of the telescope which 
is 0.81, 0.59, and 0.35 at 86, 230, and 340 GHz, respectively. 
Calibration uncertainties for data covering such a large observing period have been adopted 
to be 10\%, 15\%, 20\%, and 30\% at 3, 2, 1, and 0.8 mm, respectively. Additional uncertainties 
could arise from the line intensity fluctuation with time induced by the variation of 
the stellar infrared flux, which has been recently discovered by \citet{Cernicharo2014}. All 
data have been analyzed using the GILDAS package\footnote{http://www.iram.fr/IRAMFR/GILDAS}.

The data revealed several hundreds of spectral lines which could not be assigned to any 
known molecular species collected in the public CDMS \citep{Muller2005} and JPL \citep{Pickett1998} 
spectral databases and in the MADEX code \citep{Cernicharo2012}. Most of these lines show the 
characteristic U-shaped or flatted profiles with linewidths of 29 km\,s$^{-1}$. However, above 250 GHz a 
significant number of lines
are very narrow and come from the dust formation zone of CW\,Leo \citep[see, e.g.,][]{Patel2011,
Cernicharo2013}. Among the unidentified lines observed with the IRAM 30-m telescope, we have 
been able to assign 112 to the rotational spectrum of SiCSi through an iterative procedure
described below and a close synergy between molecular spectroscopy and astrophysics. Some selected 
lines among those observed in IRC\,+10216 are shown in Figure~1. Nine lines observed above 
300 GHz with the IRAM 30-m telescopes and assigned to SiCSi correspond to unidentified lines
observed with the SMA by \citet{Patel2011}. The spatial distribution of some of these lines, as 
derived with the SMA, is shown in Figure~2. 

\begin{figure}
\begin{center}
\includegraphics[angle=0,width=0.9\columnwidth,angle=-90]{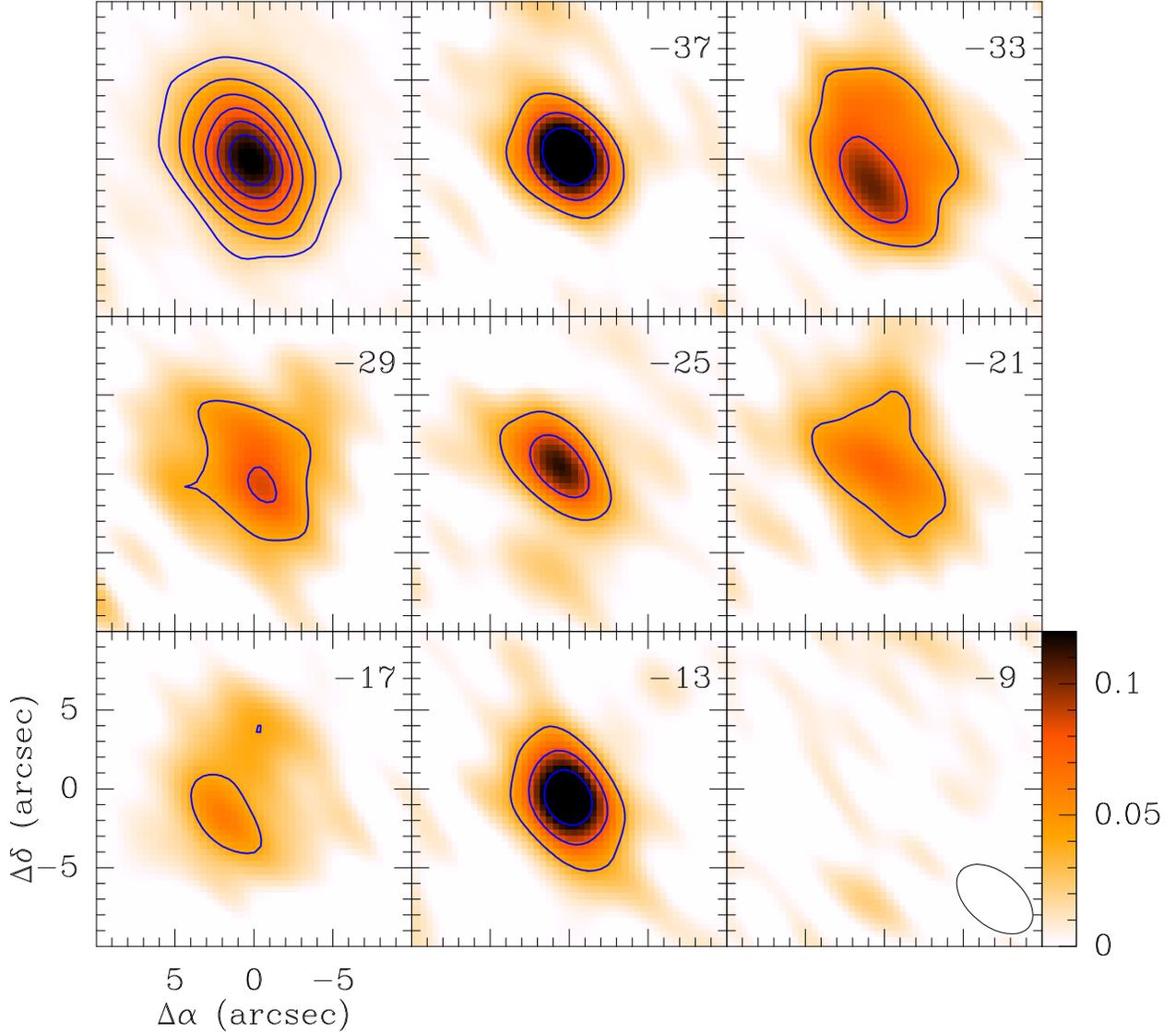}
\caption{
Maps of the average of four SiCSi lines: $18_{3,15}-18_{2,16}$, $16_{3,13}-16_{2,14}$, 
$14_{3,11}-14_{2,12}$ and $12_{3,9}-12_{2,10}$ (from the SMA line survey of IRC\,+10216 \citet{Patel2011}). 
Their visibilities were combined before 
imaging. The upper left panel is the integrated intensity map, while the remaining ones are the channel 
maps with $V_{\rm{LSR}}$ velocities indicated on top right  
in km s$^{-1}$. The size of the synthesized beam is shown in bottom right panel. 
The halftone image in the first panel is scaled linearly from 0 to 2.7 Jy beam$^{-1}$ km s$^{-1}$ 
and the contour levels (in the same units) have the starting value and intervals of 0.4 in the first panel 
and 0.04 in remaining panels.
}
\label{Fig2}
\end{center}
\end{figure}

\section{Spectroscopic constants of SiCSi}
Disilicon carbide (SiCSi) has a $C_{2v}$ symmetry and a $^1A_1$ electronic ground state with 
a modest permanent dipole moment of $\sim$1~D along the $b$-inertial axis \citep{Gabriel1992,
Barone1992,Bolton1992,Spielfiedel1996,McCarthy2015}. 
Because the two equivalent off-axis silicon atoms are bosons, only half of the rotational levels 
exist ($K_a+K_c$ even).

The search for lines of SiCSi in space began with the lines measured in the laboratory by \citet{McCarthy2015}, 
which allowed us to accurately predict frequencies with $K_a=0,1$ in the 3 mm domain. Although the molecule is 
predicted to be rather floppy, with deviations up to 10-20~MHz for  transitions higher in $J$ than those measured 
in the laboratory ($J \le10$), we quickly found 3 lines in the 3 mm spectrum of IRC\,+10216 that could be analyzed 
simultaneously with the laboratory lines. From the newly derived spectroscopic constants, which now include 
additional centrifugal distortion terms, we searched for additional lines until we found all the 
intense lines predicted in the 3 mm domain 
with upper level energies below 100~K. Using the spectroscopic constants derived from a combined fit to the laboratory and 
IRC\,+10216 data, we continued the search for $K_a=0,1$ lines in the 2 mm and 1 mm domains, and up to 20 additional 
lines were easily identified. Differences between the predicted and observed frequencies were again 10-50 MHz, but 
because the lines are so intense (typically 10-20~mK) and have a distinctive line shape, it was not difficult to 
make additional  assignments (see Figure~1).  Nevertheless, on occasion, we had to discard several initial 
assignments because subsequent predictions failed to predict new lines. With the laboratory lines and the $\sim$30 
astronomical lines involving $K_a=0$ and 1 levels, a refined set of constants were derived, and we then 
proceeded to search for lines with $K_a=2,3$ in the entire 30~m data set of IRC\,+10216 that was available to us. 
With these new astronomical measurements in hand, several additional lines with frequencies up to 180 GHz were 
measured in the laboratory. Most importantly, several of these lines correspond to transitions involving 
$K_a=1~{\rm{and}}~2$ levels, which enabled us to precisely determine the $D_K$ term, and confirm the astronomical 
assignments with higher $K_a$.

\begin{table}
\begin{center}
\caption{Spectroscopic constants (MHz) of SiCSi in comparison to SiC$_2$}
\begin{tabular}{|l|c|c||c|}
\hline
{\textbf{Constant}} & 
{\textbf{S-reduction}} &  
{\textbf{A-Reduction}} & 
{\textbf{SiC$_2$ A-reduction}} \\
\hline
$A$                            & 64074.3366(44)       & 64074.33623(37)   & 52474.1930(677)\\  
$B$                            &  4395.51772(41)      &  4395.621072(844) & 13158.71537(204)\\  
$C$                            &  4102.13098(62)      &  4102.02789(107)  & 10441.58029(168)\\  
\textit{$\Delta_J$} $\times 10^{3}$     &    9.66776(224)      &     9.73150(178)  &    13.20278(397)\\ 
$\Delta_{JK}$                &   -0.856833(73)      &    -0.8572075(610)&     1.5382238(882)\\ 
$\Delta_K $                  &   23.58788(178)      &    23.58805(148)  &    -1.2159(174)\\  
$d_1/\delta_J \times 10^{3}$ &   -1.52630(34)       &     1.519832(437) &     2.41191(253)\\  
$d_2/\delta_K \times 10^{2}$ &   -0.00318(32)       &     5.1591(454)   &     8.70880(415)\\ 
$H_J  \times 10^{8}$         &   -3.627(116)        &    -4.1349(949)   &    -8.312(460)\\  
$H_{JK} \times 10^{5}$       &    1.9373(67)        &     1.93298(547)  &    -5.0158(993)\\  
$H_{KJ} \times 10^{3}$       &    -1.8882(95)       &    -1.88755(791)  &     0.39025(399)\\  
$H_K  \times 10^{2}$         &     4.8632(204)      &     4.8633(168)   &     0.2487(529)\\  
$h_1$/$h_J \times 10^{9}$    &     -5.304(225)      &    -5.231(187)    &    -3.46(271)\\  
$h_2$/$h_{JK} \times 10^{9}$ &     -2.614(255)      &    -6586(361)     &  -35151(191)\\
$h_3$/$h_K  \times 10^3$     &                      &                    &     1.1162   (286)\\
\hline
\end{tabular}
\tablecomments{ 1 $\sigma$ uncertainties (in parentheses) are in the units of the last significant digits.}
\label{rot_const}
\end{center}
\end{table}
\normalsize

Ultimately, we were able to assign 112~lines in IRC\,+10216 to SiCSi with $J \le 48$ and $K_a\le 5$.  
At least 30\% are free of blending, and the uncertainties of the derived frequencies are less than 1~MHz. 
Other lines were slightly blended, but from their characteristic line profile it was still possible to 
measure their frequencies with an accuracy of $\sim$1~MHz. Above 250~GHz, most observations have a spectral 
resolution of 2~MHz and the frequencies are accurate to 1-1.5~MHz.  
From 
the unblended lines, an average expansion velocity of 14.0$\pm$1.0 km\,s$^{-1}$ was derived, which is 
similar to that obtained for most lines in IRC\,+10216 \citep{Cernicharo2000}. Hence, the discovery of 
SiCSi in IRC\,+10216 --which is based on 112 observed lines-- is one of the most robust identifications ever 
published for a first detection of a molecule in space. Some 20 additional lines have been also identified, 
but they are so heavily blended that accurate frequencies and intensities could not be derived.

The rotational constants were derived with a Watson Hamiltonian that included distortion constants up to 
sixth order in the $A$ and $S$ reductions (Table 1). The fit with the $S$ reduction was done with the SPFIT 
program \citet{Pickett1991}, while that in the $A$ reduction was done with fitting programs associated with 
MADEX \citep{Cernicharo2012}. We confirmed that both SPFIT and MADEX yielded the same results in the $A$ 
reduction. The rotational constants in the $A$ and $S$ reductions agree rather well, but  the $S$ reduction 
is preferred because $\delta_K$ and $h_{JK}$ are three orders of magnitude larger 
than the corresponding constants $d_1$ and $h_2$ and the predicted frequencies for high $J$ or $K_a$ 
will be less accurate. 
Because there are only a few observed lines with $K_a>3$, 
further improvement of the rotational constants was not possible. We estimate that the calculated frequencies 
with $J\le60$, $K_a\le5$ are reliable up to 400~GHz. The list of observed lines in IRC\,+10216, 
together with 
the derived line parameters, energies of the upper levels, and line strengths is provided 
in Table~2 (online), 
which also gives the frequencies and uncertainties for the 22 lines observed in the laboratory \citep{McCarthy2015}.

The SiCSi distortion constants are as large as those of SiC$_2$, a 
well-studied molecule also considered to be very floppy. Table 1 provides, for comparison purposes, the 
rotational constants derived from a fit to the lines of SiC$_2$ reported by \citet{Muller2012}
including the same constants as for SiCSi plus additional higher distortion constants. Both, $\Delta_K$ and 
$H_K$ are one order 
of magnitude larger for SiCSi than for SiC$_2$, while $\Delta_{JK} ~{\rm{and}}~\delta_K$ are of the same 
order despite SiCSi being somewhat heavier than SiC$_2$. Both molecules have a very low-lying vibrational mode, 
around 100-200~cm$^{-1}$. Hence, by analogy with SiC$_2$ for which the
antisymmetric $\nu_3$  
mode is readily observed in IRC\,+10216, we might also expect to observe rotational lines of SiCSi in 
its symmetric mode $\nu_2$. Nevertheless, SiCSi has its dipole moment along the $b$-axis while in SiC$_2$ it
is aligned along the $a$-axis which produce a significantly different rotational spectrum.

\begin{table*}
\tiny
\begin{center}
\caption{Observed lines of SiCSi in the laboratory and in IRC\,+10216}
\begin{tabular}{|rrrrrr|r|c|r|r|r|r|}
\hline
($J$ &$K_a$&$K_c$)$_u$ & ($J$& $K_a$ &$K_c$)$_l$ & $\nu_{obs}$(unc)& $\int$T$_A^* \times dv$  &Predicted$_{Lab+Astro}$(unc)&E$_{upp}$&S$_{ul}$& $\sigma$\\  
   &     &           &   &       &           &       MHz       &   K km\,s$^{-1}$         &     MHz                    &   K     &       &   mK    \\   
\hline
 5& 1& 5& 6 & 0& 6&   6718.9927 (0.002)&           &   6718.994 (0.001)&   8.9&  2.586& lab \\
 8& 0& 8& 7 & 1& 7&  12018.3548 (0.002)&           &  12018.353 (0.001)&  14.7&  3.692& lab \\
20& 1&19&19 & 2&18&  19095.4772 (0.002)&           &  19095.477 (0.002)&  89.8&  4.991& lab \\
15& 2&14&16 & 1&15&  24506.4805 (0.002)&           &  24506.481 (0.002)&  60.4&  3.644& lab \\
 3& 1& 3& 4 & 0& 4&  24959.6919 (0.002)&           &  24959.692 (0.001)&   5.3&  1.528& lab \\
10& 0&10& 9 & 1& 9&  31198.7591 (0.002)&           &  31198.761 (0.002)&  22.4&  4.861& lab \\
 1& 1& 1& 2 & 0& 2&  42663.0910 (0.002)&           &  42663.088 (0.002)&   3.3&  0.504& lab \\
 2& 1& 1& 2 & 0& 2&  60248.2200 (0.020)&           &  60248.209 (0.003)&   4.1&  2.493& lab \\
 4& 1& 3& 4 & 0& 4&  61298.7200 (0.020)&           &  61298.746 (0.003)&   7.0&  4.450& lab \\
 6& 1& 5& 6 & 0& 6&  62974.9200 (0.020)&           &  62974.923 (0.005)&  11.6&  6.340& lab \\
 8& 1& 7& 8 & 0& 8&  65310.0300 (0.020)&           &  65310.040 (0.008)&  17.8&  8.135& lab \\
 1& 1& 1& 0 & 0& 0&  68154.6000 (0.020)&           &  68154.606 (0.004)&   3.3&  1.000& lab \\
10& 1& 9&10 & 0&10&  68348.4600 (0.020)&           &  68348.474 (0.011)&  25.7&  9.806& lab \\
 3& 1& 3& 2 & 0& 2&  84424.5800 (0.020)&           &  84424.602 (0.006)&   5.3&  2.002& lab \\
 5& 1& 5& 4 & 0& 4& 100120.6600 (0.040)&           & 100120.678 (0.009)&   8.9&  3.022& lab \\
 8& 2& 6& 9 & 1& 9& 109709.6000 (0.040)&           & 109709.645 (0.018)&  26.2&  1.447& lab \\
 7& 1& 7& 6 & 0& 6& 115262.8500 (0.040)&           & 115262.899 (0.013)&  14.1&  4.073& lab \\
11& 1&11&10 & 0&10& 144033.4200 (0.040)&           & 144033.475 (0.023)&  29.3&  6.327& lab \\
 3& 2& 2& 3 & 1& 3& 180036.9000 (0.040)&           & 180036.916 (0.017)&  13.9&  1.456& lab \\
 5& 2& 4& 5 & 1& 5& 181405.4500 (0.040)&           & 181405.432 (0.014)&  17.6&  2.548& lab \\
 7& 2& 6& 7 & 1& 7& 183384.5800 (0.040)&           & 183384.603 (0.012)&  22.9&  3.561& lab \\
 9& 2& 8& 9 & 1& 9& 185976.7800 (0.040)&           & 185976.772 (0.012)&  29.8&  4.527& lab \\
16& 1&15&16 & 0&16&  82260.5000 (0.500)& 0.109 (22)&  82260.891 (0.029)&  59.2& 13.826& 1.7 \\
 3& 1& 3& 2 & 0& 2&  84424.1900 (0.330)& 0.107 (18)&  84424.602 (0.005)&   5.3&  2.002& 1.4 \\
 9& 2& 8&10 & 1& 9&  86429.8400 (0.500)& 0.059 (10)&  86429.537 (0.008)&  29.8&  1.888& 1.1 \\
18& 1&17&18 & 0&18&  88716.0400 (1.150)& 0.091 (14)&  88715.413 (0.040)&  73.7& 14.774& 1.7 \\
\hline
\end{tabular}
\tablecomments{ 1 $\sigma$ uncertainties (in parentheses) are in the units of the last significant digits.
The full table containing all observed lines can be found in the online version of the paper.}
\end{center}
\end{table*}
\normalsize

\section{Discussion}

The observed line profiles of SiCSi go from U-shaped, 
in the case of low excitation lines lying at high frequencies and thus observed with a smaller beam, 
to flat-topped, for high-excitation lines lying at low frequencies (see Figure~1). Such behaviour indicates that SiCSi 
is concentrated around the star but with a relatively extended distribution, as occurs in the case 
of metal-bearing species \citep{Cernicharo1987, Guelin1993}. The SMA maps, which correspond to transitions 
involving upper level energies between $\sim$60 and $\sim$100\,K, show that the emission is concentrated 
in a region of $\simeq6''$ in diameter (see Figure~2). A slightly larger brightness distribution size can be 
expected for  lines involving lower energy levels (see Figure~1).
We have checked if SiCSi could be responsible for some of 
the lines detected with ALMA in IRC\,+10216 \citep{Cernicharo2013}. Unfortunately, only three strong 
lines are within the spectral domain covered by these ALMA cycle0 observations and only one seems to 
be free of blending and is clearly detected in the central pixel. However, these observations were selected 
to trace the dust formation zone and their sensitivity is not good enough to trace the spatial distribution 
of weak lines.

\begin{figure}
\begin{center}
\includegraphics[angle=0,scale=0.50]{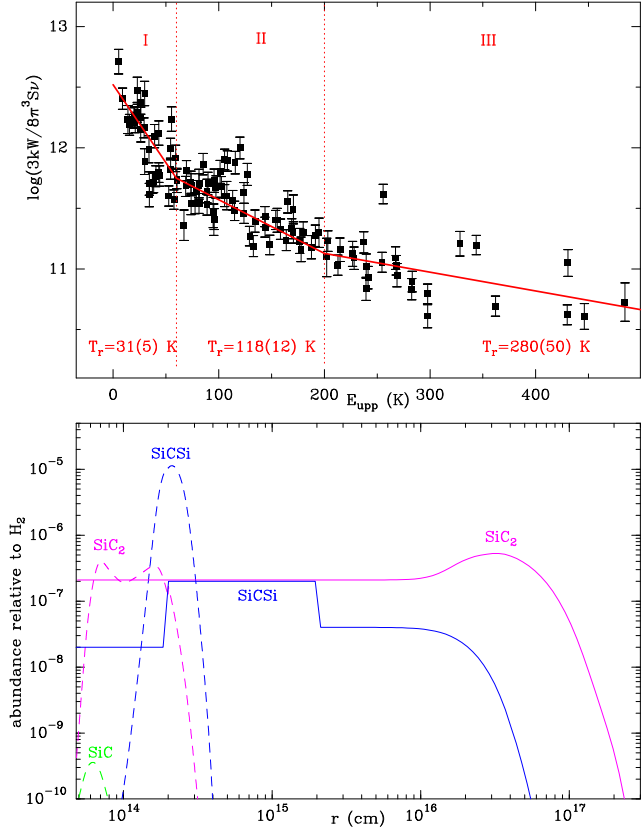}
\caption{The top panel shows the 
rotational diagram of the 112 observed lines of SiCSi. The three zones discussed in
the text are indicated by vertical dashed lines. The bottom panel shows
the radial abundance distribution of SiCSi, SiC$_2$, and SiC as calculated by chemical 
equilibrium (dashed lines) and as derived from the observations of IRC\,+10216 (solid lines). 
The derived abundance profile of SiC$_2$ is taken from \citet{Cernicharo2010}.
}
\label{Fig3}
\end{center}
\end{figure}

To initially estimate the abundance and origin of the SiCSi emission from transitions
involving different energies, we show in the top panel of Figure~3 a rotational diagram based 
on the 112 observed lines. We have adopted a gaussian source size with a radius 5$''$. The observed 
integrated 
intensities have been corrected for dilution in the beam and for the main beam efficiency of the 
telescope.
We can distinguish three zones with different rotational temperatures. The lines with upper level energies 
below 60 K (zone I) have been fit with a rotational temperature, $T_{rot}$, of $31\pm5$ K and a column density of 
$(1.4\pm0.4)\times10^{15}$ cm$^{-2}$. The lines with upper level energies between 60 and 200 K (zone II) 
have been fit with $T_{rot}=118\pm12$ K and a column density of $(3\pm0.7)\times 
10^{15}$ cm$^{-2}$. Finally, the lines with upper level energies above 200 K (zone III) have been fit with 
$T_{rot}=280\pm50$ K and a column density of $(3.2\pm0.9)\times 10^{15}$ cm$^{-2}$.
The rotational temperature derived for zone I could correspond well with the extended molecular ring observed 
at $r\simeq$14$''$ in SiC$_2$ \citep{Guelin1993}. Its value agrees well with the rotational temperature 
derived for other species \citep{Agundez2008,Cernicharo1996}. Zone II corresponds to the emission region 
of metal-bearing species and high energy lines of SiC$_2$ with $r\le5''$ \citep[Figure 2 and][]{Velilla2015}. 
Finally, the high energy lines (zone III) 
correspond to the region with $r\le2''$ traced by high$-J$ lines and vibrationally excited states 
of SiC$_2$, SiS, HCN, HNC among other species \citep{Cernicharo2010,Cernicharo2011,Velilla2015}. In this
region, \citet{Cernicharo2010} derived for SiC$_2$ $T_{rot}\simeq$204\,K and a column 
density of $\sim8\times10^{13}$ cm$^{-2}$ averaged over a beam of 30$''$. When corrected for the 
source radius of 5$''$ adopted in this work, the column density of SiC$_2$ becomes $3\times10^{15}$ 
cm$^{-2}$. Hence, SiCSi is as abundant as SiC$_2$ in the dust formation zone of IRC\,+10216. The 
large difference in the intensities of the observed lines of SiCSi and SiC$_2$, near a factor 
of 100, arise from the lower dipole moment and the larger partition function of SiCSi. 
When corrected for these effects, the observed intensities of the individual lines of SiCSi and SiC$_2$ 
indicate a SiC$_2$/SiCSi abundance ratio of  about 5.

Early chemical equilibrium calculations by \citet{Tejero1991} showed that in the inner envelope of 
C-rich AGB stars the most abundant gas phase species containing a Si$-$C bond are SiC$_2$ and SiCSi, 
while SiC is predicted to be present with a much lower abundance (see also \citealt{Yasuda2012} and 
Figure~3). To constrain the abundance and spatial distribution of SiCSi in IRC\,+10216 we have carried 
out radiative transfer calculations to obtain the emergent line profiles and compare them with the observed 
ones. The physical model of the envelope is taken from \citet{Agundez2012} with the downward revision of the 
density of particles in the regions inner to 5 R$_*$ derived by \citet{Cernicharo2013}. Since collisional 
rate coefficients are not available for SiCSi we have assumed local thermodynamic equilibrium, 
which is a reasonable approximation given the low dipole moment of SiCSi. We find that the intensities and 
profiles of the low excitation lines observed can be well reproduced by adopting a SiCSi abundance relative 
to H$_2$ of $4\times10^{-8}$, from the inner layers out to the photodissociation region. We have assumed 
that SiCSi is photodissociated by the ambient interstellar ultraviolet field with a rate of 
$10^{-9}\times\exp{(-2A_V)}$ s$^{-1}$, where $A_V$ is the visual extinction. However, to 
reproduce the high intensities observed for lines involving upper level energies above 200 K, it is 
necessary to increase the abundance of SiCSi in the warm inner regions to
$2\times10^{-7}$ (see Figure~3). Finally, a depletion in the 
very inner regions, which is consistent with the expectations from chemical equilibrium, yields a better 
agreement with the observed line profiles (see Figure~1). The derived abundance profile for SiCSi is shown 
in Figure~3. 

High angular resolution interferometric observations are needed to better constrain the 
abundance gradient of SiCSi in the inner layers. In any case, we find that disilicon carbide is as 
abundant as silicon dicarbide in the inner layers. The depletion in the abundance of SiCSi at 50 
R$_*$, which leads to a SiC$_2$/SiCSi abundance ratio of $\simeq$5 in the outer regions, would be consistent 
with its incorporation into SiC dust grains.

\acknowledgements
We thank spanish MINECO for funding 
under grants AYA2009-07304, AYA2012-32032, CSD2009-00038, and 
ERC under ERC-2013-SyG, G.A. 610256 NANOCOSMOS. The new laboratory measurements 
in Cambridge were supported by NASA Grant NNX13AE59G.


\begin{thebibliography}{}
\bibitem[Ag\'undez et al.(2007)]{Agundez2007} Ag{\'u}ndez, M., Cernicharo, J., Gu\'elin, M., 2007, \apj, 662, L91
\bibitem[Ag\'undez et al.(2008)]{Agundez2008} Ag{\'u}ndez, M., Fonfr\'ia, J. P., Cernicharo, J., et al.\ 2008, \aap, 479, 493
\bibitem[Ag\'undez et al.(2010)]{Agundez2010} Ag\'undez, M., Cernicharo, J., Gu\'elin, M., et al. 2010, \aap, 517, L2
\bibitem[Ag\'undez et al.(2012)]{Agundez2012} Ag\'undez, M., Fonfr\'ia, J. P., Cernicharo, J., et al. 2012, \aap, 543, A48
\bibitem[Ag\'undez et al.(2014)]{Agundez2014} Ag\'undez, M., Cernicharo, J., Gu\'elin, M., 2014, \aap, 570, 45
\bibitem[Barone et al.(1992)]{Barone1992} Barone, V., Jensen, P., \& Minichino, C. 1992, J. Mol. Spectrosc., 154, 252
\bibitem[Bolton et al.(1992)]{Bolton1992} Bolton, E. E., DeLeeuw, B. J., Fowler, J. E., et al. 1992, \jcp, 97, 5586
\bibitem[Cernicharo(1985)]{Cernicharo1985} Cernicharo, J., 1985, Internal IRAM report (Granada: IRAM)
\bibitem[Cernicharo \& Gu\'elin(1987)]{Cernicharo1987}Cernicharo, J., \& Gu\'elin, M. 1987, \aap, 183, L10
\bibitem[Cernicharo et al.(1989)]{Cernicharo1989} Cernicharo, J., Gottlieb, C.A., Gu\'elin, et al., 1989 \apj, 341, L25
\bibitem[Cernicharo \& Gu\'elin(1996)]{Cernicharo1996}Cernicharo, J., \& Gu\'elin, 1996a, \aap, 309, l27
\bibitem[Cernicharo et al.(2000)]{Cernicharo2000}Cernicharo, J., Gu\'elin, M., \& Kahane, C. 2000, \aaps, 142, 181
\bibitem[Cernicharo et al.(2007)]{Cernicharo2007} Cernicharo, J., Gu\'elin, M., Ag\'undez, M., et al. 2007, \aap, 467, L37
\bibitem[Cernicharo et al.(2008)]{Cernicharo2008}Cernicharo, J., Gu\'elin, M., Ag\'undez, M. et al., 2008, \apj, 688, L83
\bibitem[Cernicharo et al.(2010)]{Cernicharo2010}Cernicharo, J., Waters, L.B.F.M., Decin, L., et al., 2010, \aap, 521, L8
\bibitem[Cernicharo et al.(2011)]{Cernicharo2011} Cernicharo, J., Ag\'undez, M., Kahane, C., et al. 2011, \aap, 529, 3
\bibitem[Cernicharo(2012)]{Cernicharo2012}Cernicharo, J., 2012, in ECLA-2011: Proc. of the European Conference on Laboratory Astrophysics, 
EAS Publications Series, 2012, Editors: C. Stehl, C. Joblin, \& L. d'Hendecourt (Cambridge: Cambridge Univ. Press), 251
\bibitem[Cernicharo et al.(2013)]{Cernicharo2013}Cernicharo, J., Daniel, F., Castro-Carrizo, A., et al., 2013, \apj, 778, L25
\bibitem[Cernicharo et al.(2014)]{Cernicharo2014}Cernicharo, J., Teyssier, D., Quintana-Lacaci, G., et al., 2014, \apj, 796, L21
\bibitem[Cernicharo et al.(2015)]{Cernicharo2015}Cernicharo, J., Marcelino, N., Ag\'undez, M., Gu\'elin, M., 2015, \aap, 575, A91
\bibitem[Fonfr\'ia et al.(2006)]{Fonfria2006} Fonfr\'ia, J.P., Ag\'undenz, M., Pardo, J.R., et al., 2006, \apj, 646, L127
\bibitem[Gabriel et al.(1992)]{Gabriel1992}Gabriel, W., Chambaud, G., Rosmus, P., et al. 1992, \apj, 398, 706
\bibitem[Gail (2010)]{Gail2010}Gail, H.P., 2010, Lect. Notes Phys. 815, 61
\bibitem[Gu\'elin et al.(1993)]{Guelin1993} Gu\'elin, M., Lucas, R., Cernicharo, J. 1993, \aap, 280, L19
\bibitem[Gu\'elin et al.(1997)]{Guelin1997} Gu\'elin, M., Cernicharo, J., Travers, M. J., et al. 1997, \aap, 317, L1
\bibitem[Gu\'elin et al.(2004)]{Guelin2004} Gu\'elin, S. Muller, Cernicharo, J. et al., 2004, \aap, 426, L49
\bibitem[Matsuura et al.(2011)]{Matsuura2011} Matsuura, M., Dwek, E., Meixner, M., et al., 2011, Science, 333, 1258
\bibitem[Matsuura et al.(2015)]{Matsuura2015} Matsuura, M., Dwek, E., Barlow, M., et al., 2015, \apj, 800, 50
\bibitem[McCarthy et al.(2006)]{McCarthy2006}McCarthy, M.C., Gottlieb, C.A., Gupta, H., Thaddeus, P., 2006, \apj, 652, L141
\bibitem[McCarthy et al.(2015)]{McCarthy2015}McCarthy, M., et al., 2015, J. Phys. Chem. Lett., submitted
\bibitem[Menten et al.(2012)]{Menten2012}Menten, K.M., Reid, M.J., Kaminski, T., Claussen, M.J., 2012, \aap, 543, A73
\bibitem[M\"uller et al.(2005)]{Muller2005}M\"uller, H.S.P., Schl\"oder, F., Stutzki, J., Winnewisser, G., 2005, J. Mol. Struct., 742, 215
\bibitem[M\"uller et al.(2012)]{Muller2012}M\"uller, H.S.P., Cernicharo, J., Ag\'undez, M., et al., 2012, J. Mol. Spectrosc., 271, 50
\bibitem[Pardo et al.(2001)]{Pardo2001}Pardo, J. R., Cernicharo, J., Serabyn, E. 2001, IEEE Trans. Antennas and Propagation, 49, 12
\bibitem[Patel et al.(2011)]{Patel2011} Patel, N.A., Young, K.H., Gottlieb, C.A., et al., 2011, \apjs, 193, 17
\bibitem[Patel et al.(2013)]{Patel2013} Patel, N.A., Gottlieb, C.A., Young, K.H., 2013, in
The Life Cycle of Dust in the Universe: Observations, Theory, and Laboratory Experiments - LCDU 2013, Taipei, Taiwan
\bibitem[Pickett(1991)]{Pickett1991}Pickett, H.M., 1991, J. Mol. Spectrosc., 148, 371
\bibitem[Pickett(1998)]{Pickett1998}Pickett, H.M., Poynter, R.L., Cohen, E.A., et al., 1998, J. Quant. Spectrosc. Radiat. Transfer, 60, 883
\bibitem[Spielfiedel et al.(1996)]{Spielfiedel1996} Spielfiedel, A., Carter, S., Feautrier, N., et al., 1996, J. Phys. Chem., 100, 10055
\bibitem[Tejero \& Cernicharo(1991)]{Tejero1991}Tejero, J., and Cernicharo, J. 1991, Modelos de Equilibrio Termodina\'amico Aplicados a Envolturas Circunestelares de Estrellas Evolucionadas (Madrid: IGN)
\bibitem[Thaddeus et al.(1984)]{Thaddeus1984} Thaddeus, P., Cummins, S.E., Linke, R.A., 1984, \apj, 283, L45
\bibitem[Thaddeus et al.(2008)]{Thaddeus2008}Thaddeus, P., Gottlieb, C. A., Gupta, H., et al. 2008, \apj, 677, 1132
\bibitem[Treffers \& Cohen(1974)]{Treffers1974} Treffers, R., \& Cohen M., 1974, \apj, 188, 545
\bibitem[Velilla Prieto et al.(2015)]{Velilla2015}Velilla Prieto L., et al., 2015, submitted to ApJ Letters
\bibitem[Yasuda \& Kozasa(2012)]{Yasuda2012}Yasuda, Y. \& Kozasa, T. 2012, \apj, 745, 159
\end{thebibliography}
\end{document}